\documentclass[prd,aps,showpacs,superscriptaddress,preprint]{revtex4-1}
\pdfoutput=1

\usepackage{graphicx}

\def\lsim{\mathrel{\rlap{\lower4pt\hbox{$\sim$}}
    \raise1pt\hbox{$<$}}}                
\def\gsim{\mathrel{\rlap{\lower4pt\hbox{$\sim$}}
    \raise1pt\hbox{$>$}}}                

\begin{document}

\preprint{NSF-KITP-11-098}

\title{The QCD trace anomaly}

\author{Jens O. Andersen}
\affiliation{Department of Physics, Norwegian University of Science and Technology, N-7491 Trondheim, Norway}

\author{Lars E. Leganger}
\affiliation{Department of Physics, Norwegian University of Science and Technology, N-7491 Trondheim, Norway}

\author{Michael Strickland}
\affiliation{Department of Physics, Gettysburg College, Gettysburg, Pennsylvania 17325, USA}
\affiliation{Kavli Institute for Theoretical Physics, University of California, Santa Barbara, CA 93117}
\affiliation{Frankfurt Institute for Advanced Studies,  D-60438 Frankfurt am Main, Germany}

\author{Nan Su}
\affiliation{Faculty of Physics, Bielefeld University, D-33615 Bielefeld, Germany}
\affiliation{Department of Physics, Norwegian University of Science and Technology, N-7491 Trondheim, Norway}

\begin{abstract}
In this brief report we compare the predictions of a recent next-to-next-to-leading order 
hard-thermal-loop perturbation theory (HTLpt) calculation of the QCD trace anomaly to available
lattice data.  We focus on the trace anomaly scaled by $T^2$ in two cases:  $N_f=0$ and $N_f=3$.
When using the canonical value of $\mu = 2\pi T$ for the renormalization scale, we find that for 
Yang-Mills theory ($N_f=0$) agreement between HTLpt and lattice data for the $T^2$-scaled
trace anomaly begins at temperatures on the order of $8\,T_c$ while when including quarks ($N_f=3$) agreement
begins already at temperatures above $2\,T_c$.  In both cases we find that at very high temperatures
the $T^2$-scaled trace anomaly increases with temperature in accordance with the predictions 
of HTLpt.
\end{abstract}

\maketitle 
\newpage
\cleardoublepage


\noindent
\textsc{Introduction}~:
The QCD equation of state (EoS) is of crucial importance for a better understanding of the matter created in the ultrarelativistic heavy-ion collisions~\cite{rhicexperiment}, as well as the candidates for dark matter in cosmology~\cite{darkmatter}. Due to the nonperturbative nature of QCD at low temperatures and small chemical potentials where the system is strongly coupled, lattice gauge theory is by far the most successful method for determining the EoS in this part of the phase diagram. However, due to the asymptotic freedom of QCD, lattice calculations should be compatible with perturbation theory at very high temperatures where the coupling constant becomes asymptotically weak. 

The calculation of QCD thermodynamics utilizing weakly-coupled quantum field theory has a long history \cite{weak,bn,helsinki}. The free energy is now known up to order $g_s^6 \log g_s$ \cite{helsinki}. Unfortunately, a straightforward application of perturbation theory is of no quantitative use at the phenomenology-relevant intermediate coupling regime which is probed in the ongoing ultrarelativistic heavy-ion collision experiments at Brookhaven National Laboratory's Relativistic Heavy Ion Collider and the European Organization for Nuclear Research's Large Hadron Collider (LHC). The issue is that the weak-coupling expansion series oscillates wildly and shows no sign of convergence unless the temperature is astronomically high. For example, simply comparing the magnitude of low-order contributions to the QCD free energy with three quark flavors ($N_f=3$), one finds that the $g_s^3$ contribution is smaller than the $g_s^2$ contribution only for $g_s \lsim 0.9$ or $\alpha_s \lsim 0.07$ which corresponds to a temperature of $T \sim 10^5\,$GeV $\sim 5 \times 10^5 \, T_c$, with $T_c \sim 170\,$MeV being the QCD critical temperature. 

The poor convergence of the weak-coupling expansion of thermodynamic functions stems from the fact that at high temperature the classical solution is not described by massless particles, but instead massive quasiparticles due to plasma effects such as the screening of (chromo)electric fields and Landau damping.  One way to deal with the problem is to use an effective field theory framework in which one treats hard modes using standard four-dimensional QCD and soft modes using a dimensionally reduced three-dimensional SU(3) plus adjoint Higgs model \cite{helsinki,helsinki2,helsinki3} but treating the soft sector non-perturbatively by not expanding the soft contribution
in powers of the coupling constant \cite{Blaizot:2003iq}.  This picture of treating the soft sector non-perturbatively is ubiquitous and there
exist several ways of systematically reorganizing the perturbative series at finite temperature~\cite{reorg}.  Such treatments are based on a quasiparticle picture in which one performs a loop expansion around an ideal gas of massive quasiparticles, rather than an ideal gas of massless particles. In this brief report, we will present results for the trace anomaly in pure Yang-Mills ($N_f=0$) and QCD with $N_f=3$ obtained using the hard-thermal-loop perturbation theory (HTLpt) reorganization~\cite{htlpt,ymnnlo,qcdnnlo}.  We will focus on the quantity $({\cal E}-3{\cal P})/T^2$ which, in pure Yang-Mills, has been shown to be approximately constant for temperatures which are close to the phase transition temperature.  We will show that, when using the trace anomaly scaled by $T^2$, $({\cal E}-3{\cal P})/T^2$, as a criterion, pure Yang-Mills exhibits a transition to perturbative behavior at temperatures on the order of $8\,T_c$ while for $N_f=3$ one sees reasonable agreement with the next-to-next-to-leading order (NNLO) HTLpt result already at temperatures on the order of $2 -3\,T_c$.

\vspace{2mm}

\noindent
\textsc{Background}~: 
HTLpt is a reorganization of the perturbative series for thermal gauge theory that is a gauge-invariant generalization of screened perturbation theory (SPT) applied to scalar $\phi^4$ theory~\cite{spt}. In scalar $\phi^4$ theory the basic idea of SPT is to add and subtract a thermal mass term from the bare Lagrangian: including the added piece in the free part of the Lagrangian; while treating the subtracted piece as an interaction term on the same footing as the quartic term. In gauge theories, however, simply adding and subtracting a local mass term violates gauge invariance~\cite{gaugebreak}. Instead, one adds and subtracts the hard-thermal-loop (HTL) effective action, which dresses the propagators and vertices self-consistently so that the reorganization is manifestly gauge invariant~\cite{htl}. HTLpt thermodynamic calculations for non-Abelian theories have been recently carried out to three loops, aka NNLO~\cite{ymnnlo,qcdnnlo} and improved convergence comparing to the weak-coupling expansion is found.

At temperatures just above the phase transition, $1.1 \, T_c \lsim T \lsim 4\,T_c$, in pure Yang-Mills, lattice results find that the trace anomaly scaled by $T^2$ is approximately constant.  It has been suggested that this behavior is due to the influence of power corrections of order $T^2$~\cite{fuzzy} which are nonperturbative in nature and might be related to confinement. Phenomenological fits of lattice data which include such power corrections show that the agreement with lattice data is improved~\cite{megias}. Alternatively, the power corrections are obtained from AdS/QCD models which break conformal invariance~\cite{andreev,Gubser:2008yx,Gubser:2008ny,Noronha:2009ud}. The potential impact of the power corrections on relativistic hydrodynamics relevant to heavy ion experiments at LHC is explored in Ref.~\cite{Andreev:2011iq}.  However, it is important to mention that current AdS/QCD models 
have no firm theoretical reason for including modifications which 
induce power law corrections besides their seeming phenomenological 
importance.  In addition, the models are largely unconstrained in the 
magnitude of these corrections and the parameters in the
underlying models must be fit to lattice data.

In the last decade there have been intensive studies of the EoS on the lattice for both pure Yang-Mills and full QCD with dynamical quarks.  These studies generally measure the trace anomaly scaled by $T^4$, $({\cal E}-3{\cal P})/T^4$, and use this to determine the energy density, pressure, and entropy density of the system.  In the remainder of this brief report, we compare the NNLO HTLpt trace anomaly with lattice data from the Wuppertal-Budapest \cite{Borsanyi:2011zm,Borsanyi:2010cj}, HotQCD \cite{Bazavov:2009zn,Bazavov:2010pg}, RBC-Bielefeld \cite{Cheng:2007jq,Petreczky:2009at}, and WHOT-QCD \cite{Umeda:2008bd,Umeda:2010ye} collaborations.

\begin{figure}[t]
\begin{center}
\includegraphics[width=13cm]{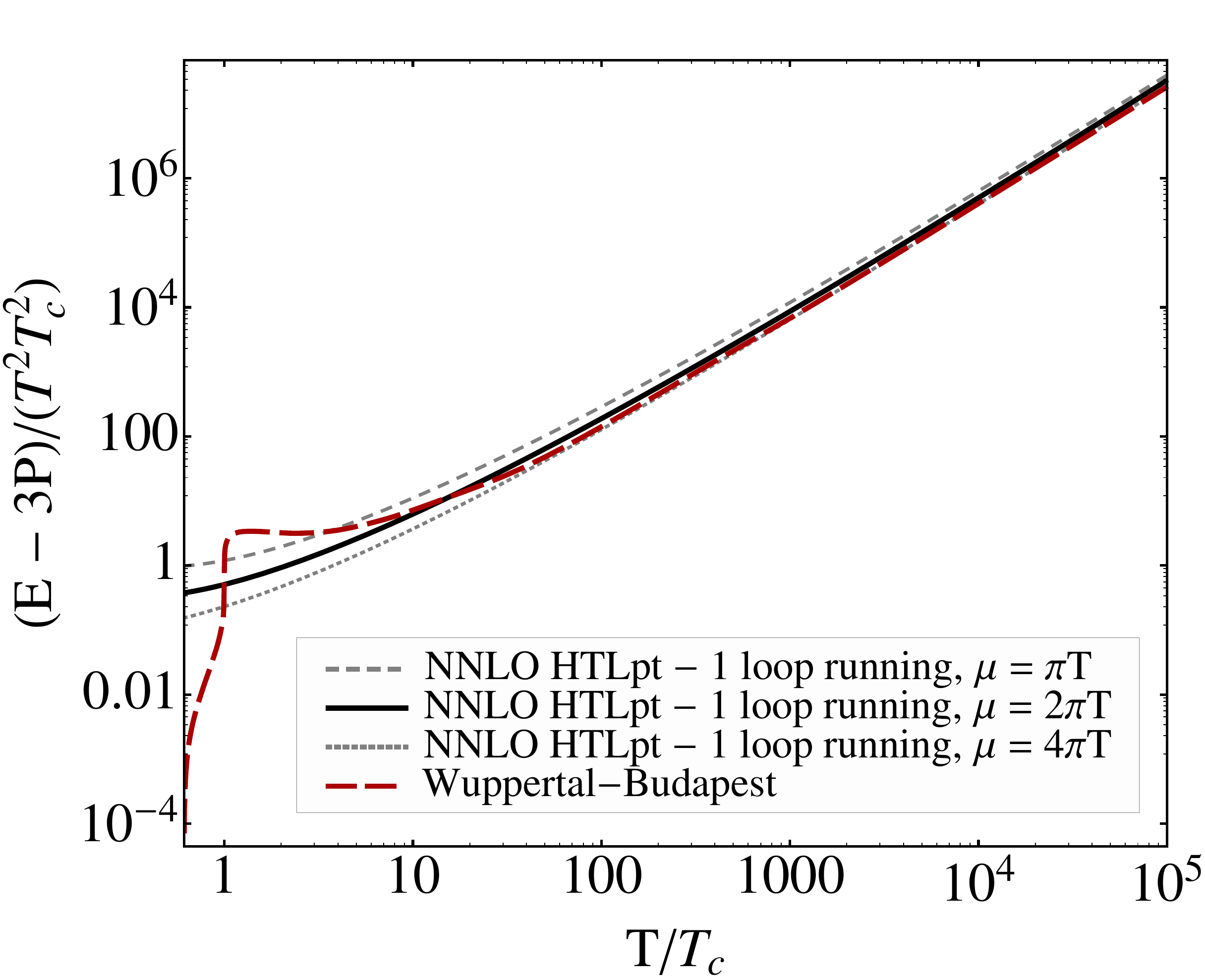}
\end{center}
\vspace{-4mm}
\caption{NNLO HTLpt result for the pure Yang-Mills ($N_f=0$) trace anomaly over temperature squared compared with lattice
data from the Wuppertal-Budapest group \cite{Borsanyi:2011zm} at high temperatures. For all curves we have
used the one-loop running of the strong coupling as specified in the text.  Different HTLpt curves correspond to
different choices of the renormalization scale $\mu$.}
\label{fig:ym1}
\end{figure}

\begin{figure}[t]
\begin{center}
\includegraphics[width=13cm]{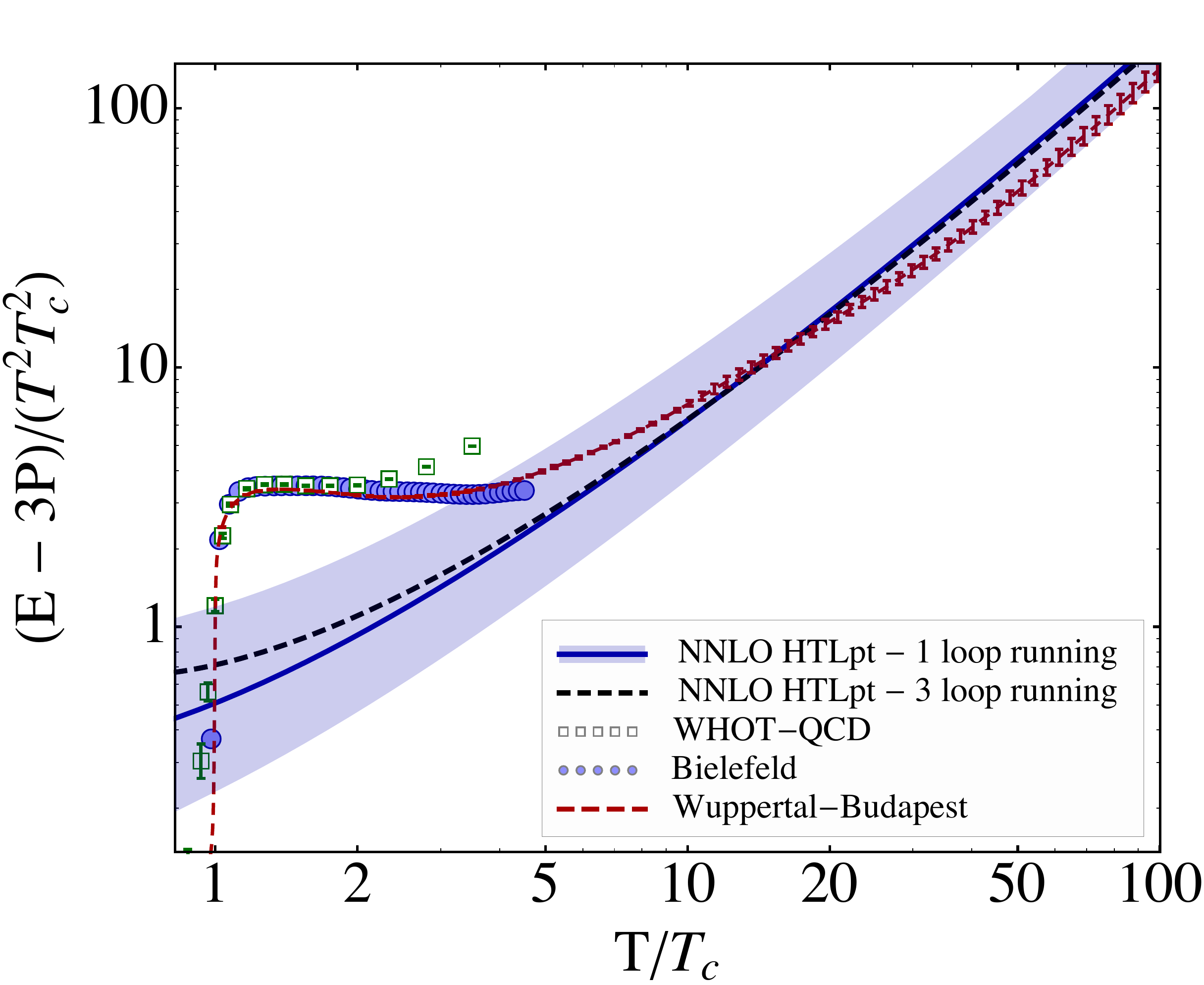}
\end{center}
\vspace{-4mm}
\caption{NNLO HTLpt result for the pure Yang-Mills ($N_f=0$) trace anomaly over temperature squared compared with lattice
data from the Wuppertal-Budapest \cite{Borsanyi:2011zm}, Bielefeld \cite{Boyd:1996bx},
and WHOT-QCD \cite{Umeda:2008bd,Umeda:2010ye} collaborations.  For the 
NNLO HTLpt result with three-loop running, the renormalization scale was taken to be $\mu = 2 \pi T$. 
For the NNLO HLTpt result with one-loop running, we show a line corresponding to  $\mu = 2 \pi T$ and
a band showing the range $\pi T \leq \mu \leq 4 \pi T$.}
\label{fig:ym2}
\end{figure}

\vspace{2mm}

\noindent
\textsc{Results}~: 
In Fig.~\ref{fig:ym1} we plot the trace anomaly scaled by $T^2T_c^2$ for pure Yang-Mills theory
and compare this with recent lattice results obtained by the Wuppertal-Budapest collaboration 
\cite{Borsanyi:2011zm}.  This figure shows the behavior of $\Delta \equiv ({\cal E}-3{\cal P})/(T^2T_c^2)$
at high temperatures.  For the NNLO HTLpt result we use the perturbative one-loop running   
of the strong coupling which is consistent with the counterterms that one obtains at NNLO.  We show
three different curves corresponding to the three different values for the renormalization scale, 
$\mu = \{\pi T, 2\pi T, 4 \pi T\}$.
The central value of $\mu = 2 \pi T$ is the scale one expects to emerge from an equilibrium calculation
since the lowest possible non-vanishing  bosonic Matsubara mode sets the scale; however, for completeness
sake we show the variation of $\mu$ by a factor of two in either direction, which should be understood 
as a dramatic overestimate of the theoretical uncertainty.
As can be seen from Fig.~\ref{fig:ym1}, at low temperatures NNLO HTLpt underpredicts the trace
anomaly for pure Yang-Mills and only starts to agree at temperatures on the order of $8\,T_c$ 
for $\mu = 2 \pi T$.  The agreement with lattice results begins at much higher temperatures 
for the other extreme values of $\mu$. 
At asymptotically high
temperatures we see excellent agreement between the NNLO HTLpt results and the lattice results.  
We note in this context that if one allows for a fit to the unknown perturbative
$g^6$ contribution to the pressure, then the EQCD approach can equally-well describe the pure Yang-Mills
lattice data down to temperatures on the order of $8\,T_c$ \cite{helsinki2,helsinki3}.
In the EQCD case, once again, the key is to not expand perturbatively in the Debye mass, but to instead treat 
contributions from this scale non-perturbatively.

The behavior of the trace anomaly at lower temperatures is presented in Fig.~\ref{fig:ym2} along with lattice data from
the Bielefeld \cite{Boyd:1996bx} and WHOT-QCD \cite{Umeda:2008bd,Umeda:2010ye} collaborations.
The solid black line is the NNLO
HTLpt result obtained using a one-loop running of the QCD coupling and the dashed black line
is the HTLpt result obtained using a three-loop running of the QCD coupling.  In the case of the 
three-loop running we use the lattice determination of $T_c/\Lambda_{\overline{\rm MS}} = 1.26$
to fix the scale \cite{Borsanyi:2011zm}.  For comparison between the one- and three-loop results we require that both
give the same value for the strong coupling when the renormalization scale $\mu = 5$ GeV.
Numerically, one finds $\alpha_s(5\;{\rm GeV}) = 0.140553$.
We compare one- and three-loop running because, formally speaking, one only obtains counterterms
consistent with one-loop running from the NNLO HTLpt calculation; however, at temperatures close
to the critical temperature the three-loop running gives important corrections to the scale dependence
of the running coupling \cite{PDG}.  We will use the difference between one- and three-loop running to gauge
the theoretical uncertainty of the NNLO HTLpt results.  For both the one- and three-loop running the lines
assume $\mu = 2 \pi T$ to fix the renormalization scale as a function of the temperature.  For the one-loop
running case we also show a band corresponding to varying the renormalization scale in the range
$\pi T \leq \mu \leq 4 \pi T$.  We note
that this value is not fit in any way to the lattice data, but instead is the predicted value of the 
appropriate renormalization scale to use.  Better agreement at high temperatures can be obtained
by fitting $\mu$ to lattice data, in which case one finds that a value of $\mu/(2\pi T) = 1.75$ is
preferred in the case of pure Yang-Mills \cite{Borsanyi:2011zm}.  Here we present the results at face
value without fitting.

The most remarkable feature of the lattice data for $\Delta$ is that it is essentially constant in the 
temperature range $T \sim 1.1 - 4\,T_c$.  At temperatures above approximately $4\,T_c$ the latest
Wuppertal-Budapest \cite{Borsanyi:2011zm,Borsanyi:2010cj} results show an upward trend in
accordance with perturbative predictions.  The WHOT-QCD results also exhibit an upward trend,
however, it starts at much lower temperatures. This discrepancy could be due to their fixed scale
approach not having sufficiently large $N_\tau$ at high temperatures as noted in their paper 
\cite{Umeda:2008bd}.

\begin{figure}[t]
\begin{center}
\includegraphics[width=13cm]{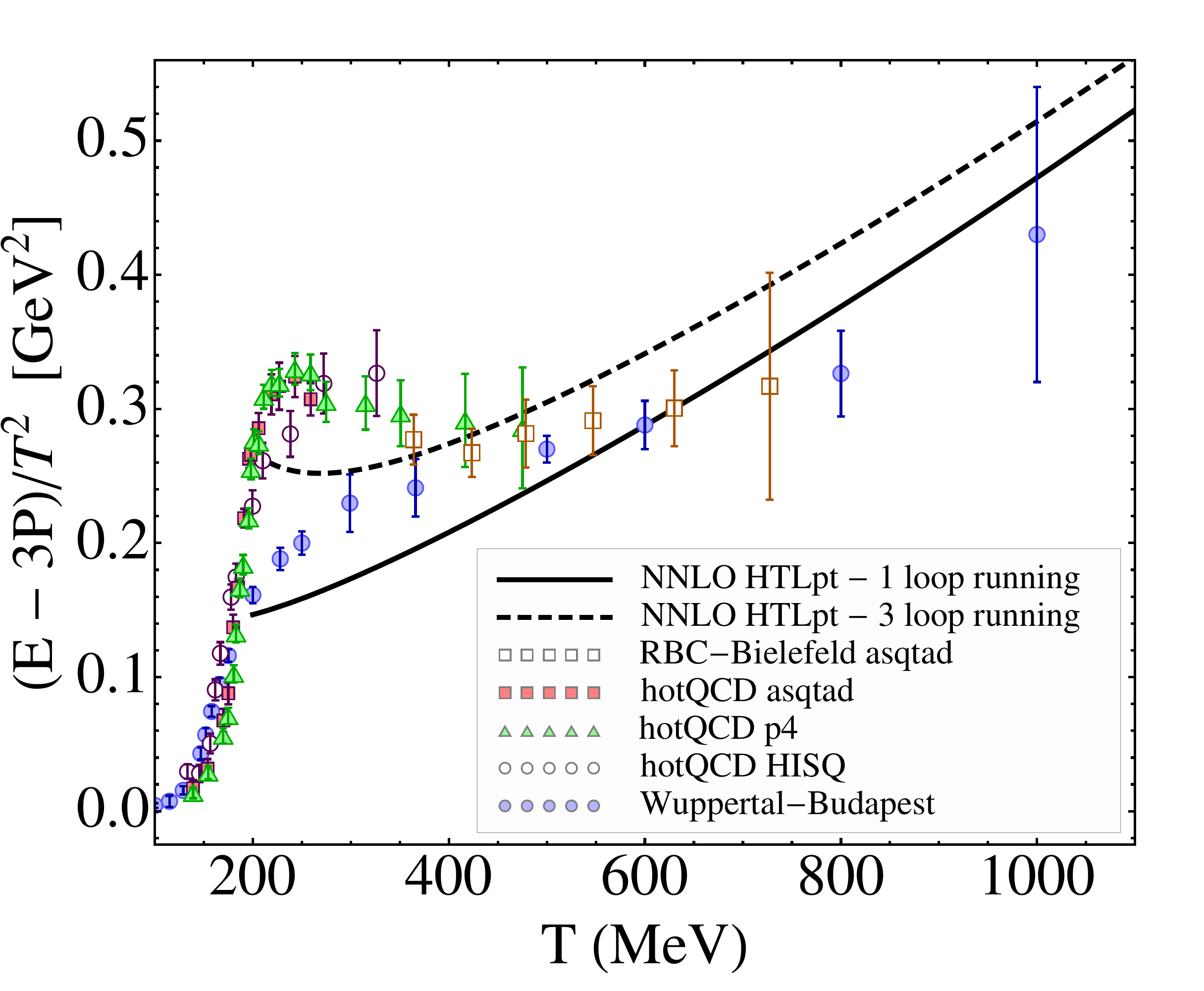}
\end{center}
\vspace{-4mm}
\caption{NNLO HTLpt result for the $N_f=3$ QCD trace anomaly over temperature squared compared with lattice
data from the Wuppertal-Budapest \cite{Borsanyi:2010cj}, hotQCD
\cite{Bazavov:2009zn,Bazavov:2010pg}, and RBC-Bielefeld \cite{Cheng:2007jq,Petreczky:2009at} collaborations.  
For both HTLpt curves the renormalization scale was taken to be $\mu = 2 \pi T$.}
\label{fig:qcd}
\end{figure}

\begin{figure}[t]
\begin{center}
\includegraphics[width=13cm]{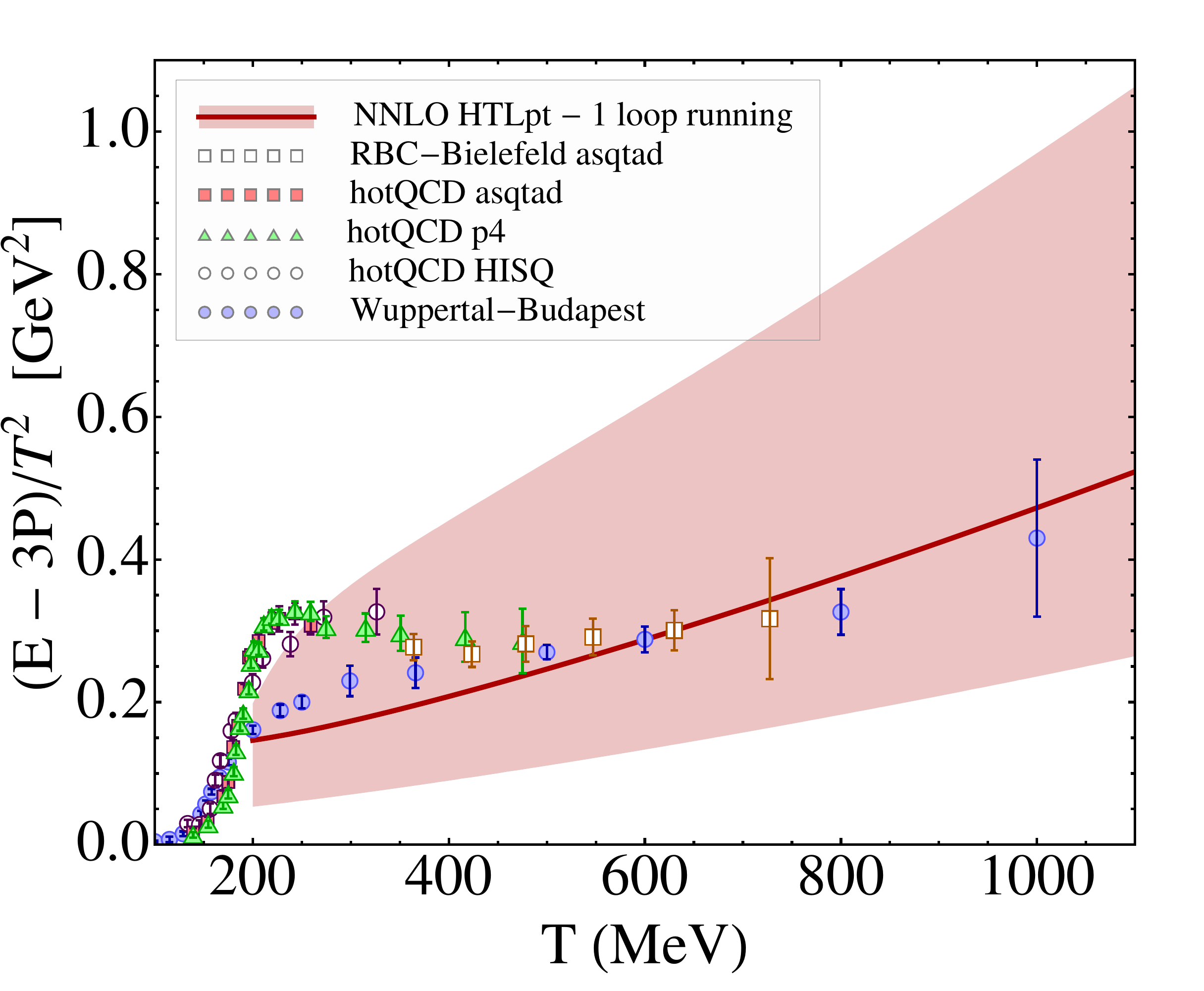}
\end{center}
\vspace{-4mm}
\caption{NNLO HTLpt result for the $N_f=3$ QCD trace anomaly over temperature squared compared with lattice
data from the Wuppertal-Budapest \cite{Borsanyi:2010cj}, hotQCD
\cite{Bazavov:2009zn,Bazavov:2010pg}, and RBC-Bielefeld \cite{Cheng:2007jq,Petreczky:2009at} collaborations.  
For the HTLpt curve we use the one-loop running of the strong coupling constant.  The red band is the
variation of the HTLpt result with the assumed renormalization group scale 
($\pi T \leq \mu \leq 4 \pi T$) and the line corresponds to the value $\mu = 2 \pi T$.}
\label{fig:qcd2}
\end{figure}

In Fig.~\ref{fig:qcd} we present the $N_f=3$ QCD NNLO HTLpt prediction for $({\cal E}-3{\cal P})/T^2$ 
and compare to lattice results available from the Wuppertal-Budapest \cite{Borsanyi:2010cj} and hotQCD
collaborations \cite{Bazavov:2009zn,Bazavov:2010pg}.  The lattice data from the Wuppertal-Budapest collaboration
uses the stout action and have been continuum estimated by averaging the trace anomaly measured using their two
smallest lattice spacings corresponding to $N_\tau = 8$ and $N_\tau = 10$.  The lattice data 
from the hotQCD collaboration are their $N_\tau = 8$ results using the asqtad, p4, and HISQ actions which
have not been continuum extrapolated \cite{Bazavov:2009zn,Bazavov:2010pg}.  
The lattice data from the RBC-Bielefeld collaboration is $N_\tau = 6$
and have also not been continuum extrapolated \cite{Cheng:2007jq,Petreczky:2009at}.

As before, we present HTLpt results using both one- and three-loop running of the strong coupling.  In this case
we have required $\alpha_s({\rm 5\;GeV}) = 0.2034$ in accordance with recent high precision lattice measurements
of the running coupling constant~\cite{McNeile:2010ji}.  As can be seen from Fig.~\ref{fig:qcd} for $T > 400$ MeV
we find reasonable agreement between the NNLO HTLpt predictions and available lattice results.  Below this
temperature the Wuppertal-Budapest and hotQCD results do not seem to agree.  Therefore, it is difficult to draw
conclusions about the efficacy of the HTLpt approach; however, naively one expects that for temperatures
less than twice the critical temperature that non-perturbative corrections should become increasingly important.

In Fig.~\ref{fig:qcd2} we again present the $N_f=3$ QCD NNLO HTLpt prediction for $({\cal E}-3{\cal P})/T^2$ 
and compare to lattice results.  However, in this case we show only the result of assuming one-loop running 
for the strong coupling constant which is consistent with the counterterms necessary to renormalize HTLpt at 
NNLO.  The red band results from varying the renormalization scale $\mu$ by a factor of two around 
$\mu = 2 \pi T$ which is the theoretically favored value.  As can be seen from this figure, it would seem at 
first glance that there is still a significant theoretical uncertainty related to the choice of the renormalization scale $\mu$; 
however, $\mu$ is not a completely free parameter since its value should be related to the lowest possible momentum 
exchange, which for a plasma in equilibrium is set by the lowest non-vanishing bosonic Matsubara mode.  In this sense, the band shown
dramatically overestimates the theoretical error of the final result; however, we include it here for completeness.

\vspace{2mm}

\noindent
\textsc{Discussion and Conclusion}~:
The purpose of this brief report is to present an easily accessible comparison of predictions of the $T^2$-scaled
trace anomaly with recent NNLO calculations using HTLpt.  The details of the HTLpt framework and the full 
NNLO calculation itself can be found in Refs.~\cite{ymnnlo,qcdnnlo}.  We compared the $N_c=3$ NNLO HTLpt 
predictions with the QCD $T^2$-scaled trace anomaly in two specific cases, namely $N_f=0$ and $N_f=3$, however,
the result for general $N_c$ and $N_f$ can be found in  Refs.~\cite{ymnnlo,qcdnnlo}.  

We find that by including quarks in the calculation, agreement with lattice data is greatly improved as compared to 
the NNLO results of pure-glue QCD.  Fermions are perturbative in the sense that they decouple in the 
dimensional-reduction step of effective field theory, so we expect that including contributions from quarks gives at 
least as good agreement with the lattice calculations as compared to the pure-glue case.  However, the reasons for 
the large improvement between the HTLpt predictions and the lattice calculations is not clear to us.

The $T^2$-scaling of the trace anomaly has received much attention recently and comparisons with lattice
data provides an extremely sensitive test of the agreement between theory and numerical results.  Using
$\mu = 2 \pi T$ we find 
that for Yang-Mills theory ($N_f=0$) agreement between HTLpt and lattice data for the trace 
anomaly begins at temperatures on the order of $8\,T_c$ while when including quarks ($N_f=3$) the agreement
begins already at temperatures above $2\,T_c$.  In both cases we find that at very high temperatures
the $T^2$-scaled trace anomaly increases with temperature in accordance with the predictions 
of HTLpt.

In closing we emphasize again that the $T^2$-scaled trace anomaly
is a very sensitive observable.  If one were to compare instead the pressure, energy density, and entropy density
one finds that, in both cases presented here, HTLpt is consistent with lattice data down to temperatures on the
order of $2-3\,T_c$ \cite{ymnnlo,qcdnnlo}.  Looking forward, we point out that the HTLpt framework is formulated in Minkowski
space and is therefore also applicable to the calculation of real-time properties of the quark gluon plasma at LHC temperatures.

\vspace{3mm}
\noindent
\textsc{Acknowledgments}~:
The authors would like to thank A. Dumitru for encouraging us to summarize our results in this form. We thank 
the Wuppertal-Budapest, hotQCD, and RBC-Bielefeld collaborations for providing their lattice data. 
M. S. was supported in part by the Helmholtz 
International Center for FAIR Landesoffensive zur Entwicklung Wissenschaftlich-\"Okonomischer Exzellenz program
and the Kavli Institute for Theoretical Physics grant No. NSF PHY05-51164. N. S. acknowledges support from the 
Yggdrasil mobility program of the Norwegian Research Council, the Sofja Kovalevskaja program of the Alexander von 
Humboldt Foundation, and the kind hospitality of the Department of Physics at the Norwegian University of Science 
and Technology.

\bibliography{anomaly}

\end{document}